# Bussgang revisited: effect of quantization on signal to distortion plus noise ratio with non-Gaussian signals

*Alister Burr, Abigail Elcock, Junbo Zhao*

*Abstract*

Quantization plays an important role in the physical layer (PHY) disaggregation which is fundamental to the Open Radio Access Network (O-RAN) architecture, since digitized signals must be transmitted over fronthaul connections.  In this paper we explore the effect of quantization on PHY performance, drawing on the Bussgang decomposition and the implications of the Bussgang theorem and extending it to the case of non-Gaussian signals.  We first prove several theorems regarding the signal to distortion plus noise ratio for a general non-linearity, applicable to both the Gaussian and the non-Gaussian case, showing that the decomposition can be applied to the non-Gaussian case, but that formulae previously introduced should be amended.  We then apply these results to the non-linearity created by quantization, both for Gaussian and non-Gaussian signal distributions, and give numerical results derived from both theory and simulation.

*Introduction*

The Open Radio Access Network (O-RAN) architecture has recently become a significant area of development: this architecture involves the disaggregation of the physical layer (PHY), with signals from the lower part of the PHY, in the radio unit (RU), being digitised and transmitted via a fronthaul connection to the distributed unit (DU) where the upper PHY is implemented.  The digitisation of course implies quantization, which introduces distortion in addition to the noise from the front end of the RU.  There is a trade-off between the effect of this distortion and the load in the fronthaul that should be analysed: the object of this paper is to explore how the Bussgang decomposition of the nonlinearity presented by the quantization can be used to model its effect.

The Bussgang decomposition arises from the Bussgang theorem of stochastic analysis, articulated by Julian Bussgang in 1952 [1].  It is worth noting that whereas the Bussgang theorem as stated in [1] assumes that the input has the Gaussian distortion, the Bussgang decomposition could with necessary adjustments be applied with an input having any given distribution.  Much more recently Zillman [2] used it to calculate a measure of the effect of non-linearity on signal integrity, the signal to distortion noise ratio (SDNR), and compares it with the mean square error (MSE).  The paper covers not only the noiseless case, when the input is itself the signal, but also where it contains both signal and noise.  Strictly speaking in the latter case SDNR should be redefined as "signal to distortion plus noise ratio".  It is important to note (though [2] did not explicitly state this) that the result given for SDNR in the latter case depends on the Bussgang theorem, and hence requires that both signal and noise are Gaussian.

In practice the signal component (at least) at the RU may well not have a Gaussian distribution: hence the main objective of this paper is to explore this case, and to show how the SDNR can then be evaluated by this means, even if the Bussgang theorem is not necessarily valid.  In the next section we revisit the Bussgang decomposition and the Bussgang theorem: we quote the main results from [2] and show that some depend on the validity of the theorem, and hence require adjustment for non-Gaussian signal distributions.  We then consider some specific examples of quantization with non-Gaussian signals.

*Bussgang theorem and Bussgang decomposition*

In its most general form the Bussgang theorem states that given two jointly Gaussian random variables $x$ and $z$, and a memoryless non-linear function $y = f(x)$, the product moment $\mathbb{E}[xz]$ of $z$ with the input $x$ and with the output $y$ are related by a constant $\alpha_x$ which depends only on the non-linearity and the standard deviation $\sigma_x$ of the input:

$$\mathbb{E}[yz] = \alpha_x \mathbb{E}[xz] \qquad (1)$$

where:

$$\alpha_x = \frac{\mathbb{E}[yx]}{\mathbb{E}[x^2]} = \frac{1}{\sigma_x^2}\int_{-\infty}^{\infty} xf(x)p_N(x,\sigma_x)dx \quad (2)$$

where $p_N(x,\sigma) = \frac{1}{\sigma\sqrt{2\pi}}\exp\left(-\frac{x^2}{2\sigma^2}\right)$ denotes the Gaussian (or normal) distribution with zero mean and standard deviation $\sigma$, and (2) follows by putting $z = x$ in (1). It further follows that the output $y$ of such a non-linearity can in general be expressed as the sum of a component which is proportional to the input (and thus fully correlated with it), and a component $\delta_x$ which is perfectly uncorrelated:

$$y = \alpha_x x + \delta_x \quad (3)$$

Assuming for the moment the noiseless case, where $x$ is therefore the signal, we may regard the first term as the signal (allowing for the scaling factor), while the second is a distortion: this is the Bussgang decomposition. It has the advantage that, as a result of the definition of $\alpha_x$, the distortion is uncorrelated with the signal term, and so in general will behave like noise. Hence we can define the SDNR as:

$$\text{SDNR} = \frac{\alpha_x^2 \mathbb{E}[x^2]}{\mathbb{E}[\delta_x^2]} \quad (4)$$

Because of the reliance of the Bussgang theorem on the Gaussian assumption, it is sometimes assumed that the Bussgang decomposition can only be used for Gaussian inputs, but this is not necessarily the case, provided the formula (2) for $\alpha_x$ is replaced by the more general expression:

$$\alpha_x = \frac{1}{\mathbb{E}[x^2]}\int_{-\infty}^{\infty} xf(x)p_x(x)dx \quad (5)$$

where $p_x(x)$ denotes the actual input distribution. [2] also defines a second parameter (there denoted by $\beta$ but here by $\gamma_x$), again only dependent on the non-linearity and the input distribution, this time relating the output power to the input power:

$$\gamma_x = \frac{\mathbb{E}[y^2]}{\mathbb{E}[x^2]} = \frac{1}{\mathbb{E}[x^2]}\int_{-\infty}^{\infty} f^2(x)p_x(x)dx \quad (6)$$

Using this, we may establish Theorem 1 (due to Zillman [2] but here explicitly not dependent on the Gaussian assumption):

*Theorem 1:* If the input to a non-linear function $f(x)$ is noiseless, and $\alpha_x$ and $\gamma_x$ are defined as in (5) and (6) respectively, then the signal to distortion noise ratio (SDNR) is given by:

$$\text{SDNR} = \frac{1}{\frac{\gamma_x}{\alpha_x^2} - 1}$$

*Proof* (also in [2]): The mean square of the distortion $\delta_x$ as defined in (3) is given by:

$$\mathbb{E}[\delta_x^2] = \mathbb{E}[(y - \alpha_x x)^2] = \mathbb{E}[y^2 + \alpha_x^2 x^2 - 2\alpha_x xy] = \mathbb{E}[y^2] + \alpha_x^2 \mathbb{E}[x^2] - 2\alpha_x \mathbb{E}[xy] \quad (7)$$
$$= (\gamma_x - \alpha_x^2)\mathbb{E}[x^2]$$

Hence:

$$\text{SDNR} = \frac{\alpha_x^2 \mathbb{E}[x^2]}{(\gamma_x - \alpha_x^2)\mathbb{E}[x^2]} = \frac{1}{\frac{\gamma_x}{\alpha_x^2} - 1} \quad (8)$$

This result follows directly from the definitions above, and does not rely on the Bussgang theorem, and hence also does not depend on the Gaussian assumption. ∎

However consider now the case where the input includes a signal plus unwanted noise/interference: we will refer to this as the noisy case. We write:

$$x = s + n \quad (9)$$

where $s$ is the signal component, with distribution $p_s(s)$ and $n$ is Gaussian noise with standard deviation $\sigma_n$. Then we can use the Bussgang decomposition to again split the output into a component which is proportional to the signal component of the input, and a component which is uncorrelated to it, writing:

$$y = \alpha(s+n) + \delta \tag{10}$$

In this case the signal component of the output is $\alpha s$, and hence we require that:

$$\mathbb{E}[(y - \alpha s)s] = \mathbb{E}[ys] - \alpha \mathbb{E}[s^2] = 0$$

$$\alpha = \frac{\mathbb{E}[ys]}{\mathbb{E}[s^2]} \tag{11}$$

For this case we will distinguish this coefficient from $\alpha_x$ as defined in (2) by writing it as $\alpha_s$. However it is of interest to consider the relationship between $\alpha_s$ and $\alpha_x$. We may write:

$$\alpha_x = \frac{\mathbb{E}[yx]}{\mathbb{E}[x^2]} = \frac{\mathbb{E}[y(s+n)]}{\mathbb{E}[(s+n)^2]} = \frac{\mathbb{E}[ys] + \mathbb{E}[yn]}{\mathbb{E}[s^2] + \mathbb{E}[n^2]}$$

$$\alpha_s = \frac{\mathbb{E}[ys]}{\mathbb{E}[s^2]} = \frac{\alpha_x(\mathbb{E}[s^2] + \mathbb{E}[n^2])}{\mathbb{E}[s^2]} - \frac{\mathbb{E}[yn]}{\mathbb{E}[s^2]} = \alpha_x\left(1 + \frac{\sigma_n^2}{\sigma_s^2}\right) - \frac{\mathbb{E}[yn]}{\sigma_s^2} \tag{12}$$

For the case where the signal component of the input is Gaussian (which we will call the Gaussian case), we can prove the following lemma:

*Lemma 1:* For the Gaussian case $\alpha_s = \alpha_x$.

*Proof:* In the Gaussian case we may use the Bussgang theorem, putting $z = n$ in (1). Then:

$$\mathbb{E}[yn] = \alpha_x \mathbb{E}[xn] = \alpha_x \mathbb{E}[(s+n)n] = \alpha_x \mathbb{E}[n^2] = \alpha_x \sigma_n^2 \tag{13}$$

Substituting this into (12) gives:

$$\alpha_s = \alpha_x\left(1 + \frac{\sigma_n^2}{\sigma_s^2}\right) - \alpha_x \frac{\sigma_n^2}{\sigma_s^2} = \alpha_x \tag{14} \blacksquare$$

In the same way as (6), we may also define:

$$\gamma_s = \frac{\mathbb{E}[y^2]}{\mathbb{E}[s^2]} \tag{15}$$

and we may then establish, for the general case:

*Lemma 2*: Regardless of the signal distribution:

$$\gamma_s = \gamma_x\left(1 + \frac{\sigma_n^2}{\sigma_s^2}\right) \tag{16}$$

*Proof:*

$$\gamma_s = \frac{\mathbb{E}[y^2]}{\mathbb{E}[s^2]} = \frac{\gamma_x \mathbb{E}[x^2]}{\mathbb{E}[s^2]} = \frac{\gamma_x \mathbb{E}[(s+n)^2]}{\mathbb{E}[s^2]} = \frac{\gamma_x(\mathbb{E}[s^2] + \mathbb{E}[n^2])}{\mathbb{E}[s^2]} = \gamma_x\left(1 + \frac{\sigma_n^2}{\sigma_s^2}\right) \tag{17} \blacksquare$$

Hence we can obtain the following two theorems, for the general and the Gaussian noisy cases:

*Theorem 2*: In general, for the noisy case, the signal to distortion plus noise ratio is given by:

$$\text{SDNR} = \frac{1}{\frac{\gamma_s}{\alpha_s^2} - 1} = \frac{1}{\frac{\gamma_x}{\alpha_s^2}\left(1 + \frac{\sigma_n^2}{\sigma_s^2}\right) - 1}$$

*Proof*: From (10,11) the signal component in the noisy case is $\alpha_s s$. Hence the mean square error is:

$$\mathbb{E}[(y - \alpha_s s)^2] = \mathbb{E}[y^2 + \alpha_s^2 s^2 - 2\alpha_x ys] = \mathbb{E}[y^2] + \alpha_s^2 \mathbb{E}[s^2] - 2\alpha_s \mathbb{E}[ys] = (\gamma_s - \alpha_s^2)\mathbb{E}[s^2] \tag{18}$$

where the final equality follows using (11). Then:

$$\text{SDNR} = \frac{\alpha_s^2 \mathbb{E}[s^2]}{(\gamma_s - \alpha_s^2)\mathbb{E}[s^2]} = \frac{1}{\frac{\gamma_s}{\alpha_s^2} - 1} = \frac{1}{\frac{\gamma_x}{\alpha_s^2}\left(1 + \frac{\sigma_n^2}{\sigma_s^2}\right) - 1} \qquad (19)$$

where the second equality follows from Lemma 2. ∎

*Theorem 3*: In the noisy Gaussian case, the signal to distortion plus noise ratio is given by:

$$\text{SDNR} = \frac{1}{\frac{\gamma_x}{\alpha_x^2}\left(1 + \frac{\sigma_n^2}{\sigma_s^2}\right) - 1}$$

*Proof*: This follows immediately from Theorem 2 and Lemma 1. ∎

Theorem 3 is the formula obtained in [2] – which we now see depends on the Gaussian input assumption.

## Application of Bussgang decomposition to quantization

In this section we apply the results obtained above to uniform quantization, using a mid-rise quantizer and focussing in particular on the noisy non-Gaussian case. Note that according to Theorem 3 SDNR in the noisy Gaussian case is given by the values of $\alpha_x$ and $\gamma_x$ defined by (2) and (6), using the Gaussian PDF for $p_x(x)$. Formulae for these in the Gaussian case are given in [3] (arXiv version). Hence here we apply the general formulae in (11) and (15) to Theorem 2 where the non-linearity $f(x)$ is the mid-rise quantizer.

This non-linearity is defined as follows:

$$f(x) = \begin{cases} \frac{(M-1)\Delta}{2} & x > \frac{(M-2)\Delta}{2} \\ \frac{i\Delta}{2} & \frac{(i+1)\Delta}{2} \geq x > \frac{(i-1)\Delta}{2}, i = -(M-3), -(M-5) \ldots (M-3) \\ -\frac{(M-1)\Delta}{2} & x \leq \frac{-(M-2)\Delta}{2} \end{cases} \qquad (20)$$

where $M = 2^m$ denotes the number of quantization intervals and $m$ is the number of bits per quantized sample, and $\Delta$ is the quantization interval. Then:

$$\alpha_s = \frac{\mathbb{E}[ys]}{\mathbb{E}[s^2]} = \frac{1}{\sigma_s^2} \iint_{-\infty}^{\infty} ys\, p_{y,s}(y,s) dy\, ds = \frac{1}{\sigma_s^2} \int_{-\infty}^{\infty} s\, p_s(s) \int_{-\infty}^{\infty} y\, p_{y|s}(y|s) dy\, ds \qquad (21)$$

Now the conditional distribution of $y$ given $s$:

$$p_{y|s}(y|s) = \delta\left(y + \frac{(M-1)\Delta}{2}\right) \int_{-\infty}^{\frac{-(M-2)\Delta}{2}} p_\mathcal{N}(x - s, \sigma_n) dx$$
$$+ \sum_{i=-M+3\ldots-1,1\ldots M-3} \delta\left(y - \frac{i\Delta}{2}\right) \int_{\frac{(i-1)\Delta}{2}}^{\frac{(i+1)\Delta}{2}} p_\mathcal{N}(x - s) dx \qquad (22)$$
$$+ \delta\left(y - \frac{(M-1)\Delta}{2}\right) \int_{\frac{(M-2)\Delta}{2}}^{\infty} p_\mathcal{N}(x - s) dx$$

Writing the indefinite integral $\int p_\mathcal{N}(x - s, \sigma_n) dx = \frac{1}{2}\text{erf}\left(\frac{x-s}{\sigma_n\sqrt{2}}\right)$ as $I(x)$, we can express this as:

$$p_{y|s}(y|s) = \delta\left(y + \frac{(M-1)\Delta}{2}\right)\left(I\left(\frac{-(M-2)\Delta}{2}\right) - I(-\infty)\right)$$
$$+ \sum_{i=-M+3\ldots-1,1\ldots M-3} \delta\left(y - \frac{i\Delta}{2}\right)\left(I\left(\frac{(i+1)\Delta}{2}\right) - I\left(\frac{(i-1)\Delta}{2}\right)\right) \qquad (23)$$
$$+ \delta\left(y - \frac{(M-1)\Delta}{2}\right)\left(I(\infty) - I\left(\frac{(M-2)\Delta}{2}\right)\right)$$

$$= I(\infty)\left(\delta\left(y - \frac{(M-1)\Delta}{2}\right) + \delta\left(y + \frac{(M-1)\Delta}{2}\right)\right)$$
$$- \sum_{i=-\left(\frac{M}{2}-1\right)}^{\frac{M}{2}-1} I(i\Delta)\left(\delta\left(y - \left(i + \frac{1}{2}\right)\Delta\right) - \delta\left(y - \left(i - \frac{1}{2}\right)\Delta\right)\right)$$
$$= \frac{1}{2}\left(\delta\left(y - \frac{(M-1)\Delta}{2}\right) + \delta\left(y + \frac{(M-1)\Delta}{2}\right)\right)$$
$$+ \frac{1}{2}\sum_{i=-\left(\frac{M}{2}-1\right)}^{\frac{M}{2}-1} \mathrm{erf}\left(\frac{s - i\Delta}{\sigma_n\sqrt{2}}\right)\left(\delta\left(y - \left(i + \frac{1}{2}\right)\Delta\right) - \delta\left(y - \left(i - \frac{1}{2}\right)\Delta\right)\right)$$
(24)

using $I(\infty) = -I(-\infty) = \frac{1}{2}$ and $-\mathrm{erf}\left(\frac{x-s}{\sigma_n\sqrt{2}}\right) = \mathrm{erf}\left(\frac{s-x}{\sigma_n\sqrt{2}}\right)$. Then we can write the expected value of $y$ given $s$:

$$\mu_{y|s}(s) = \int_{-\infty}^{\infty} y\, p_{y|s}(y|s)dy = \frac{\Delta}{2}\sum_{i=-\left(\frac{M}{2}-1\right)}^{\frac{M}{2}-1} \mathrm{erf}\left(\frac{s - i\Delta}{\sigma_n\sqrt{2}}\right) \qquad (25)$$

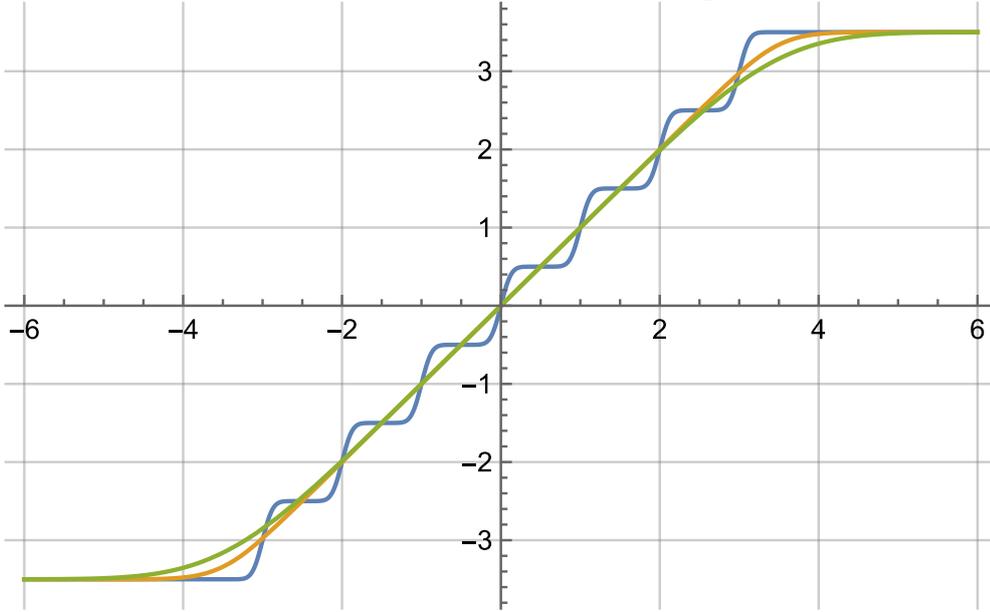

Figure 1 Plot of expected value of $y$ given $s$, plotted against $s$ for $M = 8, \Delta = 1$, and $\sigma_n = 0.1, 0.5, 0.9$ (blue, orange, green)

The result is plotted in Figure 1: we observe that, except for values of $\sigma_n$ which are very small compared to the quantization interval (and are thus unlikely to occur in practice), the resulting function is close to a straight line of gradient 1 between $\pm\frac{(M-1)\Delta}{2}$, and flat outside that range, corresponding to the truncation region.

Similarly:

$$\gamma_s = \frac{\mathbb{E}[y^2]}{\mathbb{E}[s^2]} = \frac{1}{\sigma_s^2}\iint_{-\infty}^{\infty} y^2\, p_{y,s}(y,s)dy\,ds = \frac{1}{\sigma_s^2}\int_{-\infty}^{\infty} p_s(s)\int_{-\infty}^{\infty} y^2\, p_{y|s}(y|s)dy\,ds \qquad (26)$$

Then the inner integral, which is the second moment of $y$ given $s$:

$$\mu_{y^2|s}(s) = \int_{-\infty}^{\infty} y^2 \, p_{y|s}(y|s) dy$$

$$= \frac{(M-1)^2 \Delta^2}{4} + \frac{1}{2} \sum_{i=-\left(\frac{M}{2}-1\right)}^{\frac{M}{2}-1} \text{erf}\left(\frac{s - i\Delta}{\sigma_n \sqrt{2}}\right) \left(\left(i + \frac{1}{2}\right)^2 \Delta^2 - \left(i - \frac{1}{2}\right)^2 \Delta^2\right) \quad (27)$$

$$= \frac{(M-1)^2 \Delta^2}{4} + \Delta^2 \sum_{i=-\left(\frac{M}{2}-1\right)}^{\frac{M}{2}-1} i \, \text{erf}\left(\frac{s - i\Delta}{\sigma_n \sqrt{2}}\right)$$

The final equality uses the difference of two squares formula.

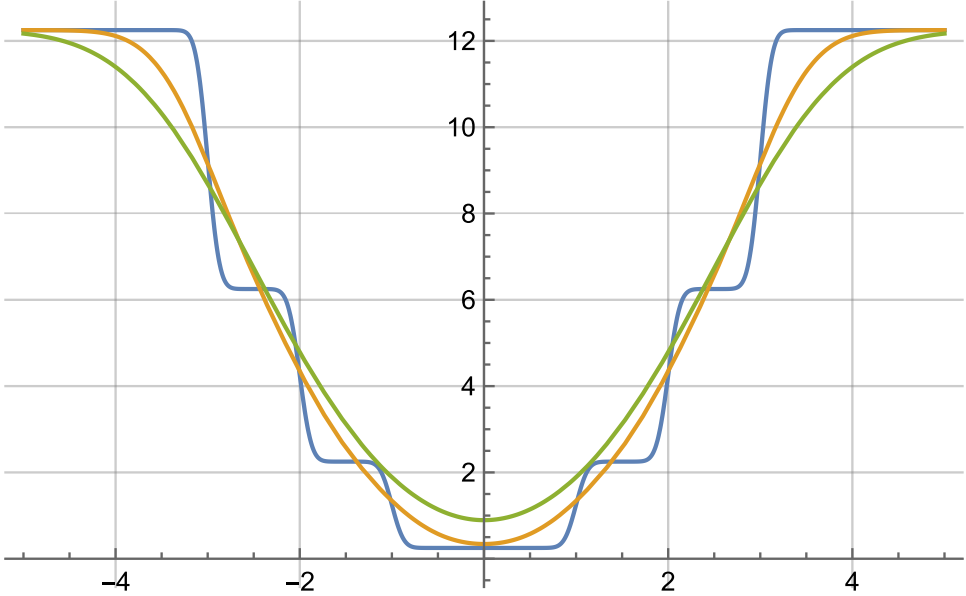

*Figure 2 Expected value of $y^2$ given s plotted against s for $M = 8, \Delta = 1$, and $\sigma_n = 0.1, 0.5, 0.9$ (blue, orange, green)*

This result is plotted in Figure 2: in this case (for moderate $\sigma_n$) the curve approximately follows a parabola between $\pm \frac{(M-1)\Delta}{2}$. The values of $\alpha_s$ and $\gamma_s$ can then be obtained from (21) and (26) respectively, and the distribution of $s$, which does not have to be Gaussian. The SDNR can then be obtained using Theorem 2.

### Numerical results and discussion

Finally we give some numerical results for two very simple cases of non-Gaussian signals: binary and 4PAM: these correspond to the distributions of in phase and quadrature samples of QPSK and 16QAM signals respectively. For the binary case we assume that $s$ takes the value $\pm A$ with equal probability; for the 4PAM case it takes values $\pm A, \pm 3A$, again with equal probability. Hence in the binary case:

$$\alpha_s = \frac{1}{\sigma_s^2} \int_{-\infty}^{\infty} s \, p_s(s) \mu_{y|s}(s) ds = \frac{\frac{1}{2} A \mu_{y|s}(A) - \frac{1}{2} A \mu_{y|s}(-A)}{A^2} = \frac{\mu_{y|s}(A)}{A} \quad (28)$$

given that $\mu_{y|s}(s)$ is an odd function of $s$. Also:

$$\gamma_s = \frac{1}{\sigma_s^2} \int_{-\infty}^{\infty} p_s(s) \mu_{y^2}(s) ds = \frac{\frac{1}{2} \mu_{y^2|s}(A) + \frac{1}{2} \mu_{y^2|s}(-A)}{A^2} = \frac{\mu_{y^2|s}(A)}{A^2} \quad (29)$$

in this case using the fact that $\mu_{y^2}(s)$ is even in $s$.

We plot results for an input signal to noise ratio (SNR) of 3 dB, by setting $A = 1$ and $\sigma_n = \sqrt{0.5}$: we plot $\alpha_s$ and $\gamma_s$, and hence the value (in linear terms, rather than dB) for SDNR.

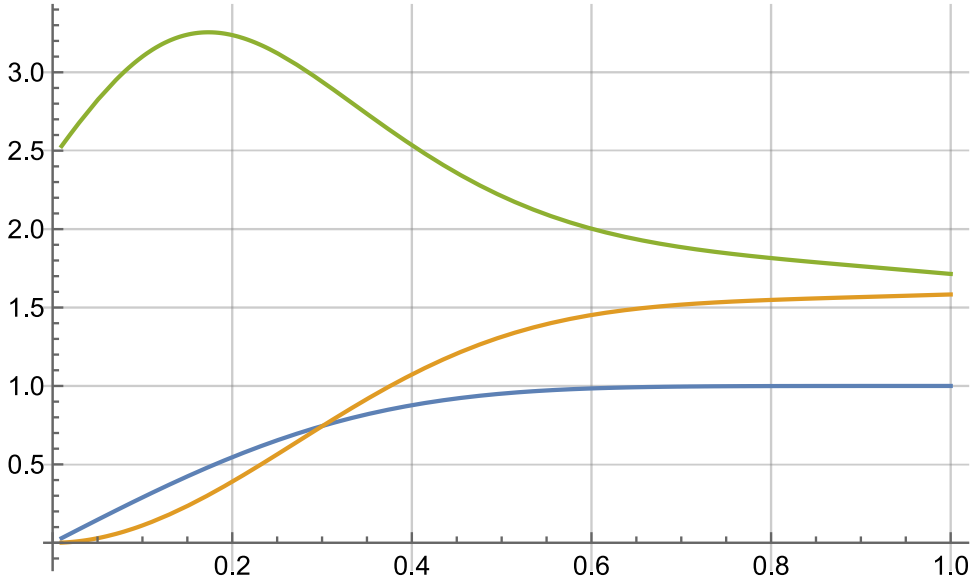

*Figure 3  Plot of $\alpha_s$ (blue), $\gamma_s$ (orange), and SDNR (green) against $\Delta$ for binary signal with A = 1, M = 8 and $\sigma_n = \sqrt{0.5}$*

Figure 3 shows the results for the binary case with $M = 8$, plotted against the quantization interval $\Delta$.  Note that for large $\Delta$, such that the quantization range is much greater than the signal amplitude $A$, $\alpha_s \to 1$, corresponding to the linear region of Figure 1.  For small $\Delta$, however, it tends to zero with $\Delta$, since the input signal is then heavily truncated.  Similarly $\gamma_s$ tends to zero with $\Delta$.  The SDNR, however, which is dependent on the ratio $\frac{\alpha_s^2}{\gamma_s}$, reaches a maximum for $\Delta = 0.175$: note that this corresponds to a truncation value of $\frac{(M-1)\Delta}{2} = 0.6125$, significantly less than the signal amplitude, which is perhaps unexpected.  Even more counterintuitively, the SDNR is then 3.63, or 5.1 dB, which is greater than the input SNR: it seems that quantization can actually improve performance.

The explanation for this is the saturation effect of the truncation, which to some degree suppresses input noise, especially for noise excursions with the same polarity as the signal.  However the distribution of the resulting distortion is far from Gaussian, which means that the bit error ratio (BER) performance is not improved.  It is easy to see that for uncoded transmission the BER is unchanged by quantization, since it depends only on the sign of the input signal, which is not affected by a mid-rise quantizer.

For the coded case it is useful to compare these results with those from [4]: this shows that quantization results in a small performance loss in terms of block error ratio (BLER) in the case of QPSK, which is equivalent to the binary case.  Further information theoretic analysis in that paper also shows a small performance loss in terms of capacity, which better matches the practical performance loss for the coded system.

The optimum quantization interval obtained in that paper is also much larger than we have obtained here, and again the optimum predicted by the information theoretic analysis better predicts it.  In particular it does not suggest a truncation level below the signal amplitude.

In the 4PAM case:

$$\alpha_s = \frac{1}{\sigma_s^2} \int_{-\infty}^{\infty} s\, p_s(s) \mu_{y|s}(s) ds = \frac{\frac{1}{4} 3A \mu_{y|s}(3A) + \frac{1}{4} A \mu_{y|s}(A) - \frac{1}{4} A \mu_{y|s}(-A) - \frac{1}{4} 3A \mu_{y|s}(3A)}{5A^2}$$
$$= \frac{3\mu_{y|s}(3A) + \mu_{y|s}(A)}{10A} \qquad (30)$$

$$\gamma_s = \frac{1}{\sigma_s^2} \int_{-\infty}^{\infty} p_s(s)\mu_{y^2}(s)ds = \frac{\frac{1}{4}\mu_{y^2|s}(3A) + \frac{1}{4}\mu_{y^2|s}(A) + \frac{1}{4}\mu_{y^2}(-A) + \frac{1}{4}\mu_{y^2|s}(-3A)}{5A^2}$$
$$= \frac{\mu_{y^2|s}(3A) + \mu_{y^2|s}(A)}{10A^2}$$
(31)

Note that in this case the mean square signal amplitude is $5A^2$, and hence we choose $A = \sqrt{0.2}$ so that mean signal power is again unity. We also choose $\sigma_n = \sqrt{0.5}$, so that once again the input SNR is 3 dB.

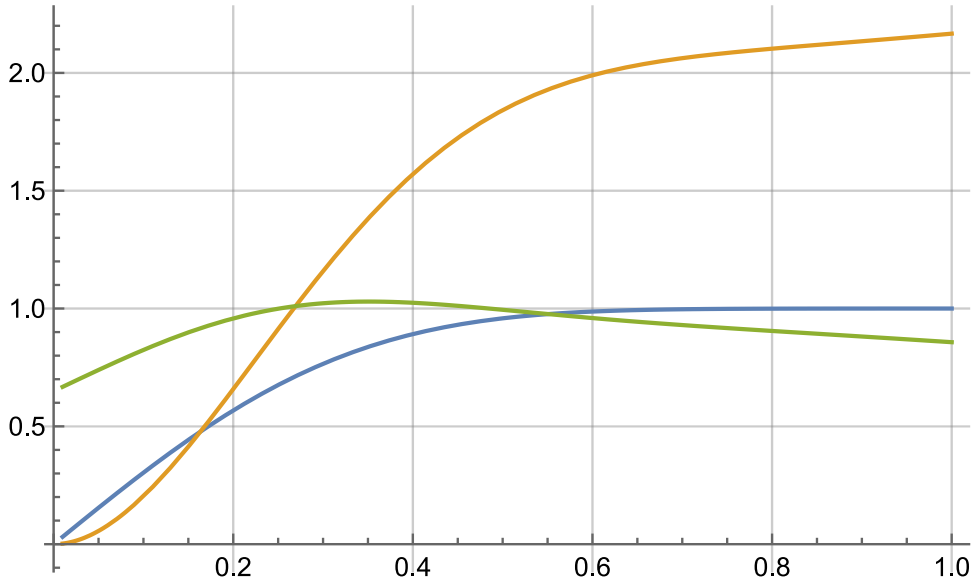

*Figure 4 Plot of $\alpha_s$ (blue), $\gamma_s$ (orange), and SDNR (green) against $\Delta$ for 4PAM signal with $A = \sqrt{0.2}$, M = 8 and $\sigma_n = \sqrt{0.5}$*

The results are plotted in Figure 4: they are similar in form, but the optimum quantization interval is larger, and the resulting SDNR is smaller: a little greater than 1, or 0 dB, so that it does not predict an SDNR greater than the input SNR. However they still do not match the results in [4] very well.

## Conclusions

We have shown that the Bussgang decomposition can be applied with non-Gaussian input signal distributions, but that the prediction of [2] for the SDNR is no longer accurate: it is shown to be dependent on the Bussgang theorem (even though that is not explicitly stated), which in turn depends on the Gaussian signal assumption. We therefore propose an alternative version in which the output of the non-linearity is split into a signal component which is correlated with the signal component of the input, and a distortion term which is fully uncorrelated with it. We derive some theorems regarding the resulting signal to distortion plus noise ratio, and the relationships between the Bussgang coefficients $\alpha_s$ and $\gamma_s$ according to this definition, and those of the original definition.

We then derive formulae for $\alpha_s$ and $\gamma_s$ applicable to the uniform quantizer for given signal distribution. These can be expressed in terms respectively of the expected values of $y$ and of $y^2$ given $s$, $\mu_{y|s}(s)$ and $\mu_{y^2|s}(s)$, and the distribution of $p_s(s)$. Interestingly these functions approximate to relatively simple forms for parameter values of interest. We then derive parameter and hence SDNR values for two simple cases, of binary and 4PAM signals, which correspond to in phase and quadrature samples of QPSK and 16QAM signals, and plot them against quantization interval $\Delta$, to determine the optimum $\Delta$ and the resulting optimum SDNR. Unfortunately these do not match very well to previous results obtained with LDPC coded systems: in particular it is possible that the resulting SDNR is greater than the input SNR to the quantizer, which is both counterintuitive and does not correspond to BER performance in practical systems. Further research is required to fully understand this result.

*References*

[1] J.J. Bussgang,"Cross-correlation function of amplitude-distorted Gaussian signals", Res. Lab. Elec., Mas. Inst. Technol., Cambridge MA, Tech. Rep. 216, March 1952

[2] P. Zillmann, "Relationship Between Two Distortion Measures for Memoryless Nonlinear Systems," in IEEE Signal Processing Letters, vol. 17, no. 11, pp. 917-920, Nov. 2010, doi: 10.1109/LSP.2010.2072498.

[3] Alister Burr, Manijeh Bashar and Dick Maryopi "Cooperative access networks: Optimum fronthaul quantization in distributed Massive MIMO and cloud RAN" 2018 IEEE 87th Vehicular Technology Conference (VTC Spring), Porto, Portugal, 2018, pp. 1-5, doi: 10.1109/VTCSpring.2018.8417560.  Also arXiv:1805.10198

[4] Alister Burr and Abigail Elcock "Optimum linear fronthaul quantization for Open-RAN systems" COST CA20120 TD(23)04009, Dubrovnik, Croatia, January 2023